\documentclass[preprint,review,12pt]{elsarticle}

\usepackage[utf8]{inputenc}
\usepackage{graphicx}
\usepackage{amsmath}
\usepackage{algorithmic}
\usepackage{subcaption}
\usepackage{relsize}
\usepackage{tikz}
\usepackage{pgfplots}
\usepackage{acronym}
\usepackage{tikz}
\usepackage{colortbl}
\usepackage{pgfplots}
\usepackage{pgfplotstable}
\usepackage{tabularx}
\usepackage{booktabs}

\usetikzlibrary{patterns,arrows,positioning,fit,shadows,backgrounds,matrix,arrows,scopes,shadows}
\usetikzlibrary{shapes.symbols,positioning}
\usetikzlibrary{intersections}
\usetikzlibrary{calc}
\makeatletter
\tikzset{nomorepostaction/.code=\let\tikz@postactions\pgfutil@empty}
\makeatother 

\definecolor{uablue50}{RGB}{0,91,130}
\definecolor{uablue35}{RGB}{166,198,211}
\definecolor{uablue30}{RGB}{178,206,217}
\definecolor{uablue20}{RGB}{204,222,236}
\definecolor{uablue10}{RGB}{229,239,242}

\definecolor{uagray80}{RGB}{81,81,81}
\definecolor{uagray50}{RGB}{156,156,156}
\definecolor{uagray30}{RGB}{185,185,185}
\definecolor{uagray20}{RGB}{204,204,204} % unofficial extension (inkscape conversion from cmyk!)
\definecolor{uagray10}{RGB}{229,229,229} % unofficial extension (inkscape conversion from cmyk!)
\definecolor{uagray05}{RGB}{242,242,242} % unofficial extension (inkscape conversion from cmyk!)
\definecolor{uawhite}{RGB}{255,255,255}

\definecolor{uabranchred}{RGB}{212,45,18}% Gesundheit
\definecolor{uabranchyellow}{RGB}{230,175,17}% Energie und Umwelt
\definecolor{uabranchblue}{RGB}{0,131,190}% Informationstechnologie

\definecolor{uablue}{RGB}{0,91,130}% Diagram color blue % 100%
\definecolor{ualightblue}{RGB}{110,159,189}% Diagram color blue % 50% (inkscape conversion from cmyk!) {cmyk}{0.42,0.16,0.00,0.26}

\definecolor{uared}{RGB}{175,90,80}% Diagram color red % 100%
\definecolor{ualightred}{RGB}{198,141,132}% Diagram color red % 70%

\definecolor{uagreen}{RGB}{125,150,110}% Diagram color green % 100%
\definecolor{ualightgreen}{RGB}{164,181,153}% Diagram color green % 70%

\definecolor{uayellow}{RGB}{215,170,80}% Diagram color yellow % 100%
\definecolor{ualightyellow}{RGB}{235,212,167}% Diagram color yellow % 50%

\acrodef{AMT}{Amazon Mechanical Turk}
\acrodef{BVLC}{Berkeley Vision and Learning Center}
\acrodef{CAD}{Computer Aided Design}
\acrodef{CDBN}{Convolutional Deep Belief Network}
\acrodef{CNN}{Convolutional Neural Network}
\acrodef{CT}{Computed Tomography}
\acrodef{CUDA}{Compute Unified Device Architecture}
\acrodef{GAN}{Generative Adversarial Network}
\acrodef{GIFT}{GPU acceleration and Inverted File Twice}
\acrodef{GPU}{Graphics Processing Unit}
\acrodef{MVCNN}{Multi-View Convolutional Neural Network}
\acrodef{OFF}{Object File Format}
\acrodef{PCD}{Point Cloud Data}
\acrodef{PCL}{Point Cloud Library}
\acrodef{PLY}{Polygon File Format}
\acrodef{SVM}{Support Vector Machine}
\acrodef{VRN}{Voxception-ResNet}
\acrodef{ReLU}{Rectified Linear Unit}

\usepackage{todonotes}
\usepackage{xcolor}

\usepackage{url}
\journal{Image and Vision Computing}

\pgfplotstableset{
    /color cells/min/.initial=0,
    /color cells/max/.initial=1000,
    /color cells/textcolor/.initial=,
    color cells/.code={%
        \pgfqkeys{/color cells}{#1}%
        \pgfkeysalso{%
            postproc cell content/.code={%
                \begingroup
                \pgfkeysgetvalue{/pgfplots/table/@preprocessed cell content}\value
                \ifx\value\empty
                    \endgroup
                \else
                \pgfmathfloatparsenumber{\value}%
                \pgfmathfloattofixed{\pgfmathresult}%
                \let\value=\pgfmathresult
                \pgfplotscolormapaccess
                    [\pgfkeysvalueof{/color cells/min}:\pgfkeysvalueof{/color cells/max}]
                    {\value}
                    {\pgfkeysvalueof{/pgfplots/colormap name}}%
                \pgfkeysgetvalue{/pgfplots/table/@cell content}\typesetvalue
                \pgfkeysgetvalue{/color cells/textcolor}\textcolorvalue
                \toks0=\expandafter{\typesetvalue}%
                \xdef\temp{%
                    \noexpand\pgfkeysalso{%
                        @cell content={%
                            \noexpand\cellcolor[rgb]{\pgfmathresult}%
                            \noexpand\definecolor{mapped color}{rgb}{\pgfmathresult}%
                            \ifx\textcolorvalue\empty
                            \else
                                \noexpand\color{\textcolorvalue}%
                            \fi
                            \the\toks0 %
                        }%
                    }%
                }%
                \endgroup
                \temp
                \fi
            }%
        }%
    }
}
\begin{document}
\begin{frontmatter}
\title{Large-scale Multiview 3D Hand Pose Dataset}

\author{Francisco Gomez-Donoso and Sergio Orts-Escolano and Miguel Cazorla}
\address{University Institute for Computer Research. University of Alicante}

\begin{abstract}
Accurate hand pose estimation at joint level has several uses on human-robot interaction, user interfacing and virtual reality applications. However, it is a currently unresolved question. The novel deep learning techniques could make great improvement in this respect but they need an enormous amount of annotated data. The hand pose datasets released so far are impossible to use in deep learning methods as they present issues such as the limited number of samples, high-level abstraction annotations or samples consisting in depth maps.

In this work, we introduce a multiview hand pose dataset in which we provide color images of hands and different kind of annotations for each, i.e. the bounding box and the 2D and 3D location on the joints in the hand. Furthermore, we introduce a simple yet accurate deep learning architecture for real-time robust 2D hand pose estimation.
\end{abstract}
\end{frontmatter}

\section{Introduction}
\label{sec:introduction}

Hand pose estimation in 2D images is a burgeoning research line that is booming in the field of computer vision. This problem is not only about the location of the hand itself, but also the localization of the hand joints, namely, the pose of the hand. The usefulness of this identification is wide ranging, from its use for sign language recognition\cite{EXSY:EXSY12160,Cazorla2015CAE} to the identification of more complex behaviors such as hand gestures. It is also worth highlighting its importance in virtual reality and augmented reality applications. In order to interact in a natural way in these virtual worlds, the hand pose detection problem must first be tackled.
Furthermore, to achieve a seamlessly integrated experience in virtual worlds, non-intrusive methods must be utilized for the hand pose detection. A way to accomplish this task is to use computer vision algorithms that work on images to retrieve the hand pose.

It is well known that deep learning systems require a large amount of labeled data to train robust models. In most cases, manual annotation is not feasible and researchers have come up with different procedures that can automatically generate ground truth annotations. Commonly, two different approaches are used for generating large amount of annotated data. First, there are tools for generating synthetic data. For example, in computer vision it is very common to use a graphics engine to simulate a real-world camera and so acquire synthetic images. However, this approach suffers from realistic-synthetic discrepancies and the trained network is therefor unable to correctly classify real-life data. The other approach is to use ad-hoc tools/sensors that can provide us with ground truth annotations. This is usually the approach that yields best results since it is able to automatize the labeling procedure and the data that is acquired is more similar to real data. 

As the review of the state of the art reveal, the first datasets were composed of color images, but they were intended to perform static or dynamic gesture recognition, so they did not provide hand joint level annotations, but a class for each samples. Then, with the emergence of low cost range sensors, the datasets began to set aside the color information to provide depth maps instead. Nonetheless, most deep learning approaches utilize color information to tackle classification and regression problems.

This paper presents a novel pose dataset. It provides images of hands from multiple points of view. The ground truth is composed of different kinds of data. For each image the hand location is provided: its bounding box, the X and Y image coordinates of each joint position of the hand, and the X, Y and Z real-world coordinates of the position of each joint of the hand. The dataset contains images of several individuals at different instants in time. There are even samples captured under special conditions which will challenge the generalization capabilities of the algorithms that use this dataset. Training and testing splits are also provided.

The rest of the paper is organized as follows. Section \ref{sec:related_works} reviews state-of-the-art works related to the existent hand pose datasets. Next, in Section \ref{sec:capture_device} the details of the capture device we use for composing the dataset are given. Then, Section~\ref{sec:dataset_description} explains relevant details of the dataset itself. Section \ref{sec:baseline} describes a system for hand pose regression trained on this dataset to be used by other hand pose estimation based proposals as a baseline. Section \ref{sec:conclusions} presents the main conclusions of the present work and some lines of future work are suggested.

\section{Related Works}
\label{sec:related_works}

Hand pose estimation is one of the most common areas of research in human-computer interaction. Since early in the $2000's$, efforts have been made to achieve highly accurate hand gesture and pose recognition through different methods. First, traditional computer vision algorithms were applied in static color images to segment the hand from the background. Then, machine learning methods were used to tackle the problem and lastly, the most sophisticated and also accurate system have been achieved through deep learning.
Alongside these methods, different kinds of datasets emerged, intended to feed these systems with proper data. In this section, we review these datasets, and highlight their weaknesses and strengths.

As previously mentioned, early in the $2000's$ a set of two datasets for static pose detection\cite{MarcelStatic1} and dynamic gesture classification\cite{MarcelDynamic1} appeared. The static pose version dataset consists of over $3,000$ images of hands in static poses distributed in $6$ different classes. The dynamic gestures dataset is composed of $60$ sequences of hands performing a certain gesture. This dataset is distributed in $4$ classes. 

In both versions, only the class of each sample is provided, with no further annotations, and the samples are highly unbalanced, which may harm the accuracy of the learning algorithms. The resolution and quantity of the samples are also clearly insufficient by today's standards.

Much later, in $2014$, Pisharady et.al\cite{Pisharady2014} released another hand pose classification dataset. The NUS hand posture dataset consists of 10 classes of postures, 24 sample images per class, which are captured by varying the position and size of the hand within the image frame. Both greyscale and color images are available at $160\times120$ pixels. 
The hand postures are selected in such a way that the inter class variation in the appearance of the postures is minimized, which makes the recognition task challenging, but the background of the images in this dataset is uniform.

Again, this dataset does not provide annotation of the joints in the hand, but a class for each sample, so it is not intended to be used in continuous hand pose estimation at joint level. Moreover, they state the background is uniform which is an undesirable feature for learning algorithms.

The dataset for hand gesture recognition introduced in \cite{Grzejszczak2016MTA} contains gestures from Polish Sign Language and American Sign Language. In addition, some special signs were included. The database consists of three series of gestures which include the subsequent data: original RGB images at different resolutions, but with great image quality; ground truth binary skin presence masks and hand feature points location. In total, this dataset contains over $1600$ samples.

Despite its high resolution and great quality ground truth, this dataset contains insufficient number of samples. There is a further and bigger drawback: the background across the samples is always the same. This might harm the generalization capabilities of the learning system, as in a real world scenario the background is constantly changing, and clearly not always grey with a great deal of contrast between the hand and the background.

Another dataset was released by Molina et. al\cite{Molina14}. This dataset is composed of natural and synthetic depth maps. Each depth map represents a segmented hand in a certain pose. The annotations include the 2D position of each joint in the depth map coordinate frame, and the depth value in that point. They included $65$ natural samples divided in $6$ different dictionaries split  by their occlusion level, pose based, motion based, or compound content. The synthetic data was generated taking a natural seed and performing up to $200$ random rotations.

This is a high representative hand pose dataset but it has a major drawback: it only provides depth maps, which forces  whoever wants to use it to acquire a 3D sensor. The number of samples is also insufficient for deep learning algorithms and the synthetic samples do not represent the essence of the real world.

Following the depth trend spurred by the emergence of low cost range sensors, the NYU Hand Pose dataset\cite{tompson14tog} was created. This dataset contains 8252 test-set and 72757 training-set frames of captured RGBD data with ground-truth hand-pose information. For each frame, the RGBD data from 3 Kinects is provided: a frontal view and 2 side views. The training set contains samples from a single user only, whilst the test set contains samples from two users. A synthetic re-creation (rendering) of the hand pose is also provided for each view.

The main drawback of this dataset is that the RGB images provided are projections of the point clouds obtained by a 3D sensor. This causes the 2D color images to be rendered in low quality, with a lot of  unknown regions out of range of the 3D sensor. Thus, for using this dataset a 3D sensor is required as well.

Another depth-based dataset is proposed in \cite{Ioannis17}. This dataset contains sequences of $14$ hand gestures performed in two ways: using one finger and the whole hand. Each gesture is performed between $1$ and $10$ times by $28$ participants, resulting in $2800$ sequences. Sequences are labelled following their gesture, the number of fingers used, the performer and the trial. Each frame of sequences contains a depth image, the coordinates of $22$ joints both in the 2D depth image space and in the 3D world space creating a full hand skeleton. Once again, this dataset does not include the corresponding RGB images so a 3D sensor is required in order to use this dataset.

In view of this review of the state of the art, we consider  there is a need for a large scale hand pose dataset composed of color images and their corresponding 3D and 2D annotations.

\section{Capture Device}
\label{sec:capture_device}

\begin{figure}[!hbt]
  \centering
  \includegraphics[width=0.47\textwidth]{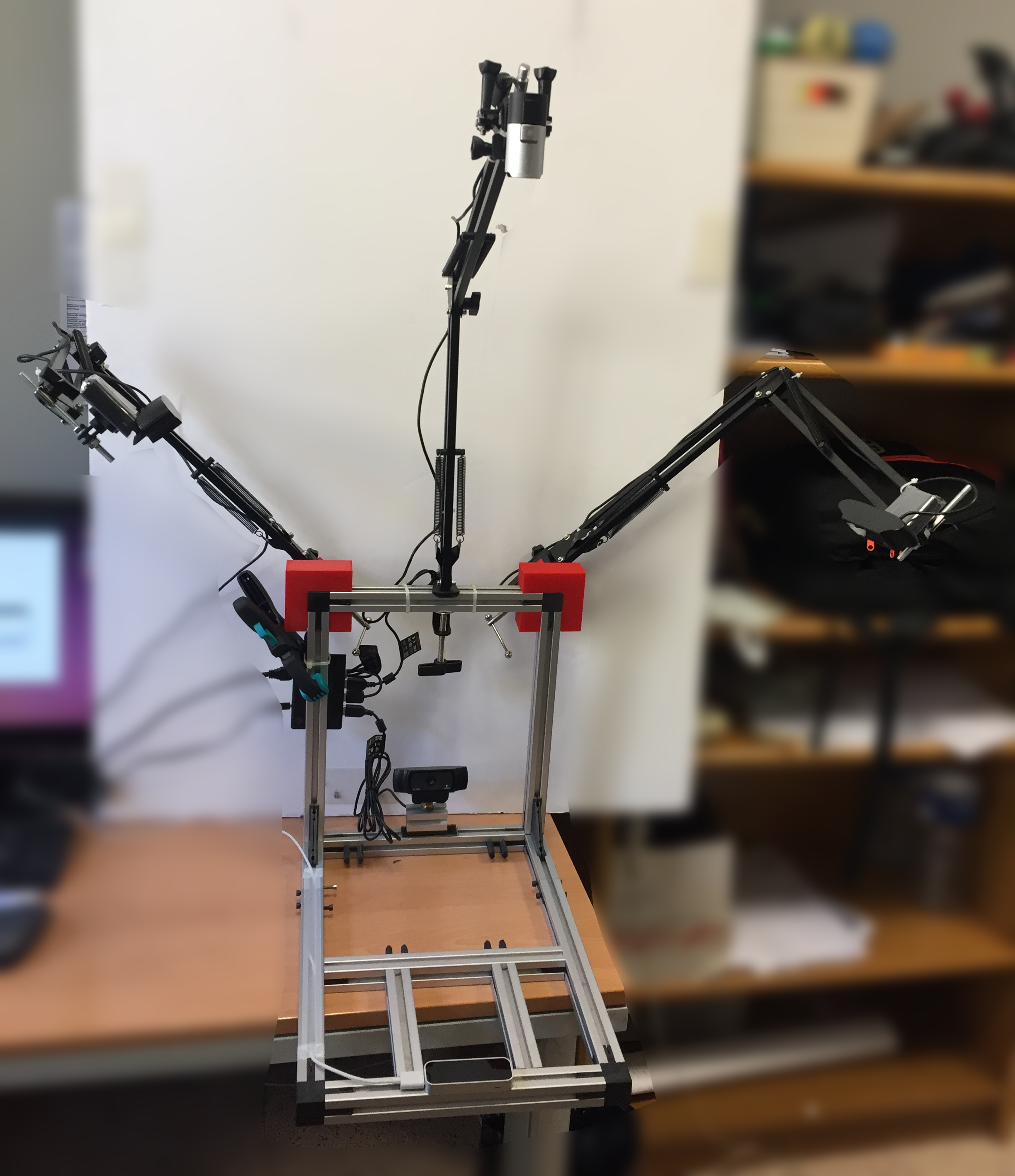}
  \caption{We created a custom structure in order to generate the dataset. This device holds the color cameras and the Leap Motion Controller, and allows us to quickly generate large amounts of ground truth data.}
  \label{fig:capture-booth}
\end{figure}

To generate accurate ground truth, a custom capture device was created, as shown in Figure \ref{fig:capture-booth}. This device consists of an aluminum structure with three articulated arms that holds a total of four cameras. Three are Logitech C920 Pro cameras, and the last one, in a zenithal position, is a Microsoft LifeCam Studio. These cameras provide high quality images up to $1080$p resolution, and although this model comes with auto focus function, this feature was disabled as it would affect the intrinsic camera parameters estimation. As the cameras are calibrated, these parameters may not vary. The images were captured at $640\times480$ resolution.

The location of the cameras is as follows: one is located in the arm located in the middle arm, facing downwards; two more are located in the side arms with an orientation of $45$ degrees; the last camera is located at the base of the structure. Finally, there is a Leap Motion Controller fixed to the aluminum frame, in front of the cameras.

The Leap Motion Controller is a device that accurately captures the pose of a person's hands. It provides sub-millimeter hand tracking at $200$Hz. This device supplies, among other data, the 3D position of the joints of a hand. This information is obtained in order to automatically annotate the position of the joints in the color images. The device is attached at the bottom of the structure, facing upwards. This configuration creates an overlapping space that is captured by both the cameras and the Leap Motion Controller. 

This device can capture an enormous amount of accurate ground truth with no effort. 
\section{Dataset Description}
\label{sec:dataset_description}

The proposed dataset is intended to be used for 3D and 2D hand pose estimation and hand area location. The dataset is structured in various sequences and each sequence is composed of a set of frames. A frame is a collection of  different kinds of ground truth data for a certain instant in time. 

Th ground truth data provided includes:
\begin{itemize}
\item A set of 3D points of the joints in the hand as provided by the Leap Motion, without further processing.
\item Four color images of a hand as captured for each color camera.
\item Four sets of 2D points as the resultant projection of the 3D points to each color camera coordinate frame.
\item Four bounding boxes computed from the projection of the 3D points to each camera coordinate frame.
\end{itemize}

Finally, it is worth noting that each sequence was captured in different conditions to assure high variability. Subject, moment of the day and speed of motion are some of the parameters we considered to achieve high variability. Further details on generalization features are given in \ref{subsec:variability}.

\subsection{3D Points of the Joints in a Hand}
\label{subsec:3d_points_joints}

For each frame a set of 3D point is provided. These 3D points correspond to the joints in the hand as provided by the Leap Motion Controller with no further preprocessing. Table \ref{tab:joint-correspondences} shows a list of each point and its semantic correspondence. In Figure~\ref{fig:calibrated}, the different points are marked on a real hand.

We provide a 3D point for each knuckle in a finger, plus the fingertips. Additionally, we also provide the palm position and the normal of the palm.

\begin{table}[!htb]
\centering
\begin{tabular}{|r|l|l|r|l|l|}
\hline
\multicolumn{1}{|c|}{Joint ID} & \multicolumn{1}{c|}{Finger} & \multicolumn{1}{c|}{Bone} & \multicolumn{1}{c|}{Joint ID} & \multicolumn{1}{c|}{Finger} & \multicolumn{1}{c|}{Bone} \\ \hline
1                                  & Index                       & Distal                    & 11                                & Pinky                       & Intermediate              \\ \hline
2                                  & Index                       & Metacarpal                & 12                                & Pinky                       & Proximal                  \\ \hline
3                                  & Index                       & Intermediate              & 13                                & Ring                        & Distal                    \\ \hline
4                                  & Index                       & Proximal                  & 14                                & Ring                        & Metacarpal                \\ \hline
5                                  & Middle                      & Distal                    & 15                                & Ring                        & Intermediate              \\ \hline
6                                  & Middle                      & Metacarpal                & 16                                & Ring                        & Proximal                  \\ \hline
7                                  & Middle                      & Intermediate              & 17                                & Thumb                       & Distal                    \\ \hline
8                                  & Middle                      & Proximal                  & 18                                & Thumb                       & Metacarpal                \\ \hline
9                                  & Pinky                       & Distal                    & 19                                & Thumb                       & Intermediate              \\ \hline
10                                 & Pinky                       & Metacarpal                & 20                                & Thumb                       & Proximal                  \\ \hline
\end{tabular}
\caption{Joints of the hand present in the dataset. The Joint Number columns reference the numerical label shown in Figure \ref{fig:calibrated}.}
\label{tab:joint-correspondences}
\end{table}

\subsection{Color Images of a Hand}
\label{subsec:color_images}

Our proposed dataset is intended to be used, but not exclusively, by deep learning models. These methods would take images of hands as input  hopefully producing accurate pose estimations. Having said that, the quality of the images is a key factor in order to make these systems work.

It is well known that images captured by a sensor reflect a slightly deformed version of the real scenario. Some of the more notorious deformations are caused by the camera's lens curvature (i.e. radial and tangential lens distortion). To deal with this problem, and to provide the best ground truth possible, an undistortioning process was carried out for each image as described in \cite{159901}. No further preprocessing was applied to the images.

As stated before, the dataset provides four color images for each frame. These four images picture a hand in the same time instant but from different perspectives.

\subsection{2D Points of the joints of a Hand}
\label{subsec:calibration}

Besides the 3D points of the joints in a hand, we also provide the 2D correspondences in each color image. In order to be able to automatically annotate images from a color camera using the Leap Motion Controller, we had to calibrate both devices, estimating the extrinsic parameters between their coordinate frames.  

The calibration process requires two steps. First, estimation of intrinsic camera parameters, $K$, of the color camera and then estimation of extrinsic parameters between the leap motion controller and the color camera, $[R | T]$. In this work, we use \cite{Zhang2000} for computing the intrinsic camera parameters. Then, we estimate the transformation that allows us to transform 3D points in the leap motion coordinate frame to the camera coordinate frame. 

\begin{figure}[!hbt]
  \centering
  \includegraphics[width=0.7\textwidth]{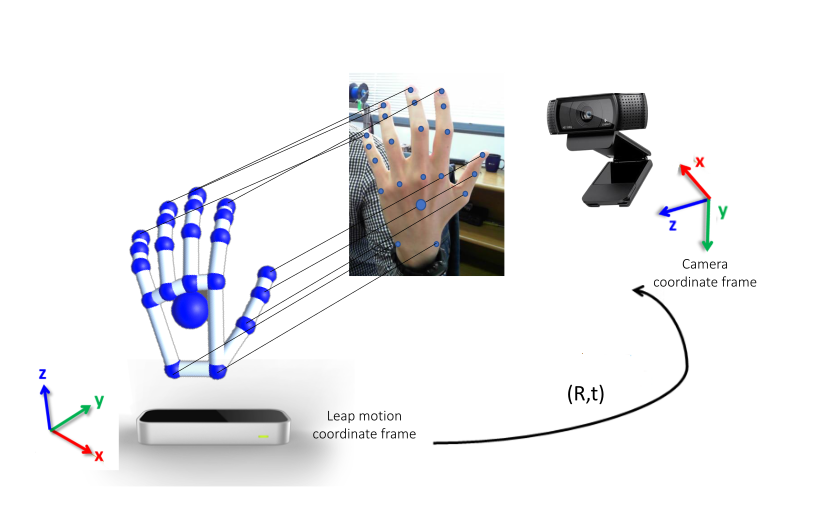}
  \caption{Calibration setup used for extrinsic parameter estimation by solving the Perspective-n-Point problem.}
  \label{fig:calib-setup}
\end{figure}

\begin{figure}[!hbt]
  \centering
  \includegraphics[width=0.47\textwidth]{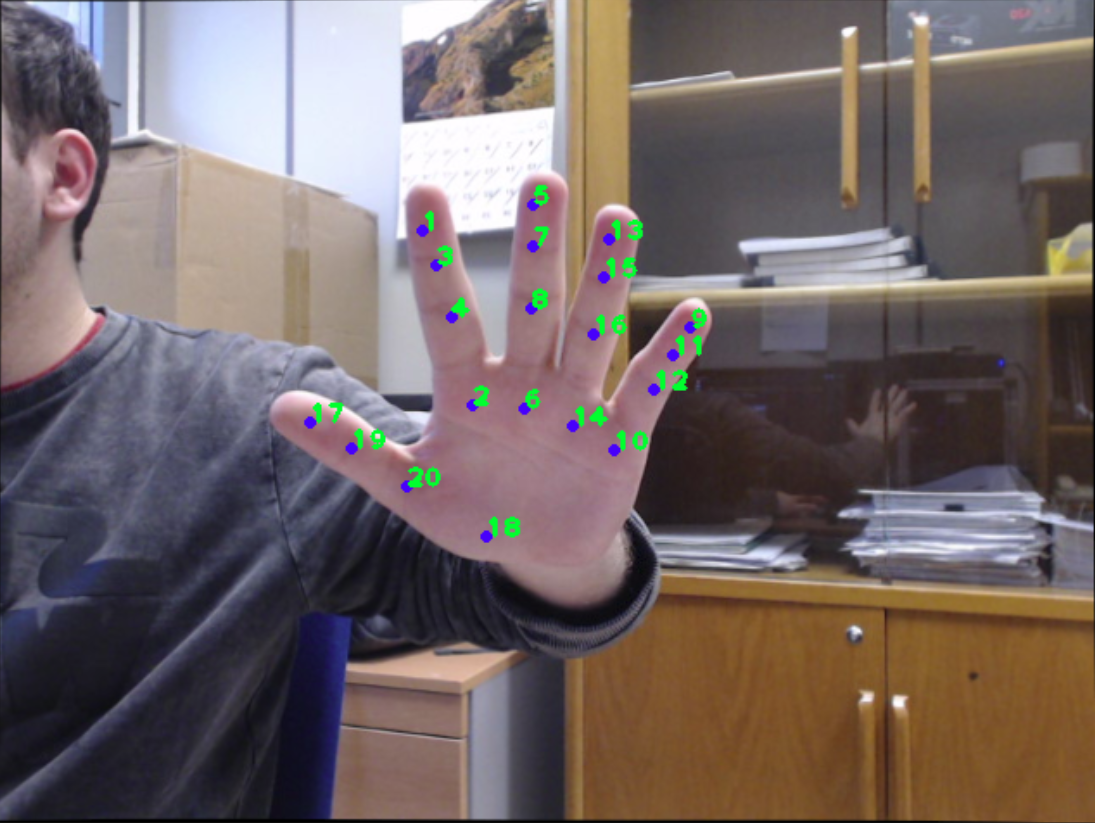}
  \caption{Once the devices are calibrated and the transformation matrix computed, new examples can be annotated in an automatic fashion in order to generate reliable ground truth.}
  \label{fig:calibrated}
\end{figure}

To estimate the transformation between these two sensors, we use a set of correspondences between leap motion data and color camera data. Given these correspondences between the 3D Leap Motion controller points and their projection on the 2D images, we captured $10$ color frames and manually labeled corresponding 3D joints positions of each fingertip. We computed the transformation matrix between the two coordinate frames by solving the Perspective-n-Point problem. Figure \ref{fig:calib-setup} illustrates this process. Basically, given a set of $n$ 3D points in the Leap motion reference frame and their corresponding 2D image projections, as well as the intrinsic camera parameters, $K$, we determine the six degree-of-freedom (6DoF) pose of the camera in the form of its rotation $R$ and translation $T$ with respect to the leap motion coordinate frame. We use the Levenberg-Marquardt optimization algorithm to minimize the final reprojection error. Figure \ref{fig:calibrated} illustrates the result of the reprojected 3D points after the calibration procedure.

As stated before, we use this transformation matrix to project the 3D points provided by the Leap Motion Controller on the color image.

As the calibration parameters are unique for each camera, we applied this method to the four cameras of the capture device. We thus, generate four sets of 2D points for each frame, as a result of projecting the 3D points provided by the Leap Motion Controller to each color camera.

\subsection{2D Bounding Box of a Hand}
\label{subsec:bbox}

Finally, we also provide the exact location of the hand in the scene for each sample through its bounding box. The bounding box is computed by taking the projection of the 3D points to the camera coordinate frame and then extracting the maximum and minimum values for the X and Y coordinates. A fixed offset of $30$ pixels is added to each direction to include the whole hand. Since there are $4$ color images per frame, the dataset provides $4$ different bounding boxes, one per color image.

\subsection{Variability and Generalization Capabilities}
\label{subsec:variability}

As previously mentioned, this dataset for pose estimation was composed to provide accurate ground truth with high variability data. To ensure this, the creation of the dataset involved $9$ different individuals recorded at different moments. The motion of the hand was spontaneous and unrestricted, with no plan at all. Therefore, a large variety of combinations are covered, in contrast to gesture-based dataset in which some fixed poses are considered. In addition, some sequences were carried out under special conditions: there are two sequences in which the subject wore different gloves, and a sequence in which the subject wore a mask. Finally, it is worth to remark that all the subjects were right-handed.

In total, we provide over $20,500$ different frames distributed in $21$ sequences. For each frame, $4$ color images and $9$ different annotations were provided, so over $80,000$ color images and over $184,500$ annotations are provided.

In order to generate the training and test splits, we shuffled all samples and took the $20\%$ of them for the test split and the remaining $80\%$ for the training split.

\begin{figure}[!hbt]
  \centering
  \includegraphics[width=0.24\textwidth, height=3.2cm]{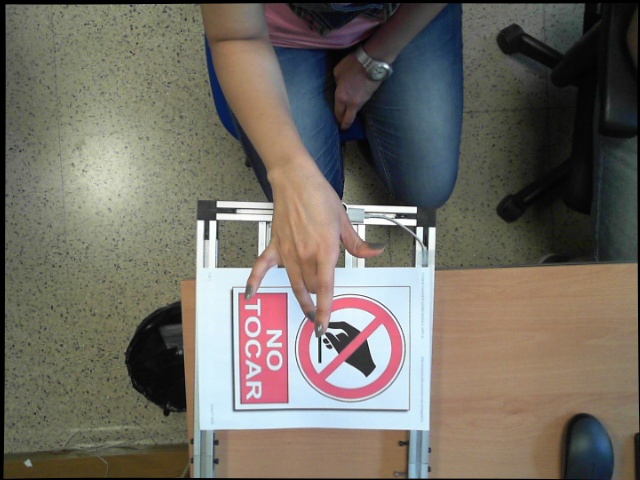}
  \vspace{1mm}
  \includegraphics[width=0.24\textwidth, height=3.2cm]{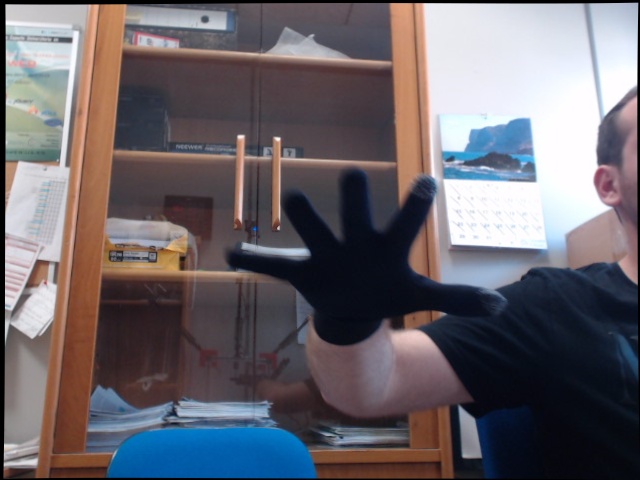}
  \includegraphics[width=0.24\textwidth, height=3.2cm]{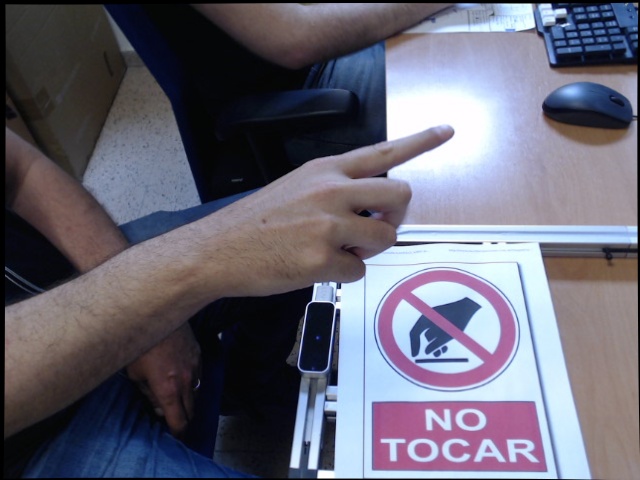}
  \vspace{1mm}
  \includegraphics[width=0.24\textwidth, height=3.2cm]{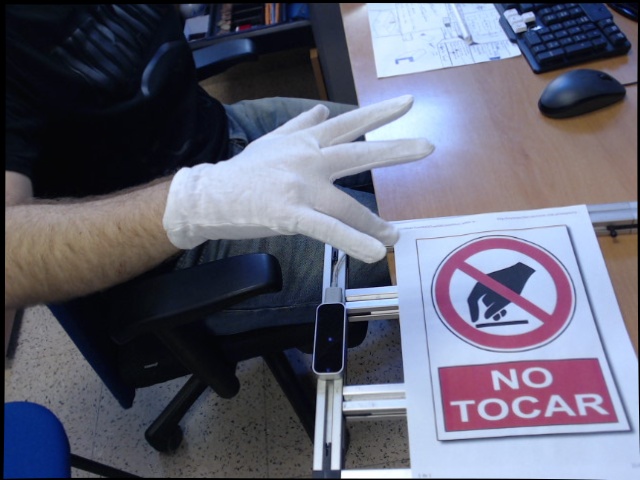}  		     \includegraphics[width=0.24\textwidth, height=3.2cm]{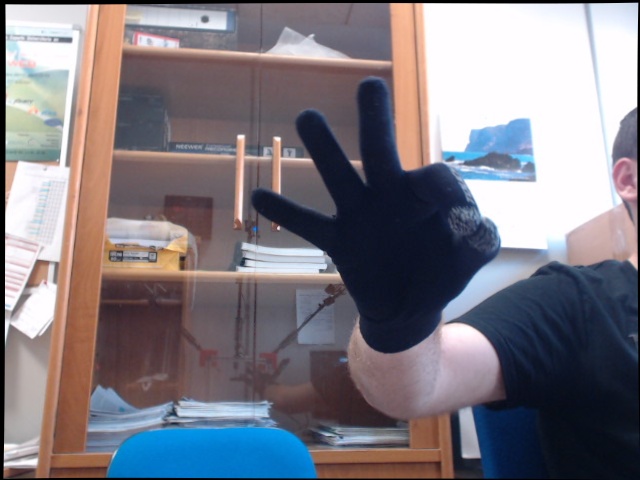}
    \includegraphics[width=0.24\textwidth, height=3.2cm]{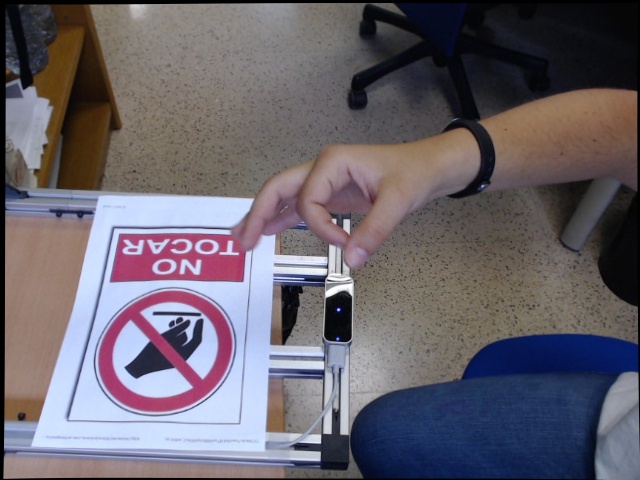}
  \includegraphics[width=0.24\textwidth, height=3.2cm]{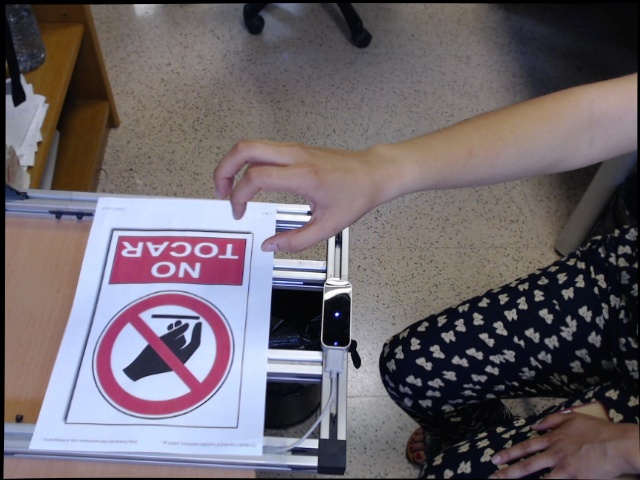}
  \includegraphics[width=0.24\textwidth, height=3.2cm]{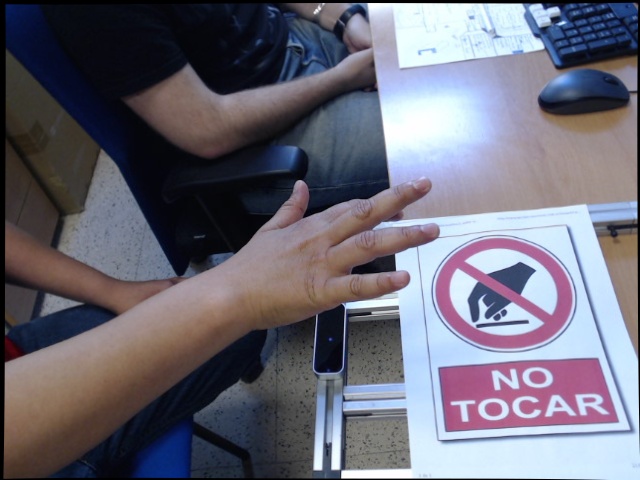}
 
  \caption{Some images extracted from the proposed dataset. Notice the high variability of the samples.}
  \label{fig:ok-poses}
\end{figure}

\section{Baseline - 2D Joints Estimation}
\label{sec:baseline}

\begin{figure*}[!hbt]
  \centering
  \includegraphics[width=0.95\textwidth]{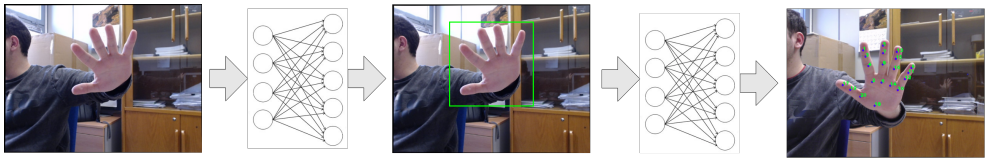}
  \caption{Overall pipeline of the proposed method. First, an image is captured using a regular web camera. This image is forwarded to a hand detector in order to obtain the bounding box of the hand. Then, the hand is cropped and passed to the hand pose regressor to finally estimate the position of the hand joints.}
  \label{fig:pipeline}
\end{figure*}

In this section, we present a baseline system for 2D hand pose estimation in color images using a deep learning approach. To do so, the following pipeline is proposed: first, a regular camera is used in order to capture a color image. This image is forwarded through a Faster R-CNN network \cite{FasterRCNN_2015} trained to detect the hands present in the scene. This network outputs both the bounding box of a hand and a score that represents the detection confidence. If this score is above $0.85$, the estimated bounding box is used to extract the hand from the image. The crop of the hand is resized to $224\times224$ pixels and fed to a custom ResNet50\cite{HeZRS15} network. By using this network in regression mode we are able to estimate $X$ and $Y$ coordinates in the image plane for each joint position. The overall pipeline is shown in Figure \ref{fig:pipeline}.

Notice that there are two critical stages in this pipeline: the hand detection (bounding box estimation) and the hand pose regression stage. These subsystems are detailed in the following subsections.

Finally, it is worth noting that this pipeline provides highly accurate hand pose estimation while maintaining a reasonable computation cost. These features allow the system to run under real-time constraints. Further results and experimentation are detailed in Section \ref{sec:experimentation}.

%% hand detection subsection
\subsection{Detecting Hand Location}
\label{sec:detecting-hand-location}
The proposed hand detection module is based on the work originally presented in \cite{FasterRCNN_2015}. We adopted the VGG-16 network \cite{Simonyan14cVGG16} pre-trained on the ImageNet dataset \cite{ImageNet2015} as the backbone network for feature extraction. In addition, we modified the number of outputs in final layers, since we are only interested in detecting hands. 

\begin{figure}[!hbt]
  \centering
  \includegraphics[width=0.32\textwidth]{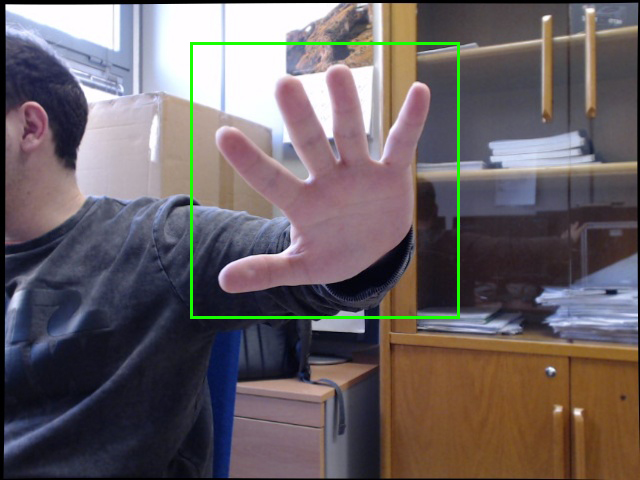}
  \includegraphics[width=0.32\textwidth]{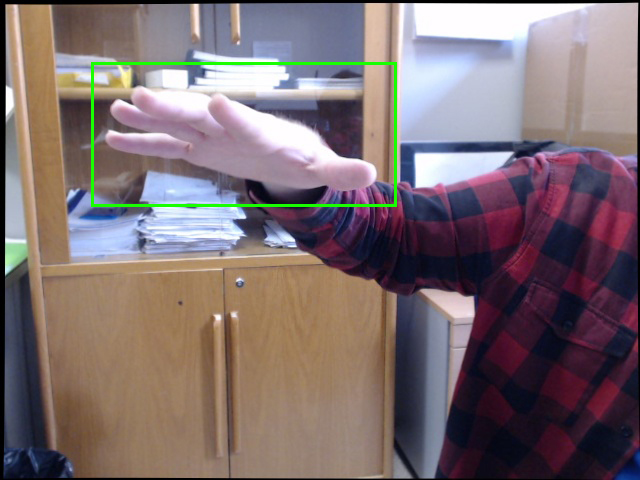}
  \includegraphics[width=0.32\textwidth]{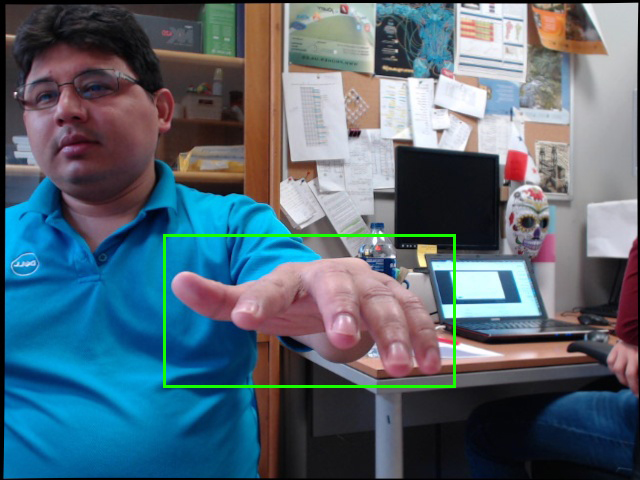}
  \caption{The Faster R-CNN network is able to accurately estimate the bounding box of a certain object, in our case, the network was trained for detecting and localizing hands.}
  \label{fig:hand-bb}
\end{figure}

We trained a Faster R-CNN network using different hands datasets. First, we used the dataset presented in \cite{DatasetHandsVGGMittal11}. This dataset contains hand images collected from various different public image data set sources as listed in \cite{DatasetHandsVGGMittal11}. We extended the number of training images (~$1000$) creating two additional sources of data. The first additional set of images was taken from the proposed dataset. The second subset of images was created using Google Images. We used the keywords 'hands' and 'close-up' to download images containing close-up hands. Finally, we also manually annotated these additional images.

The main reason to train this system on different datasets was to avoid overfitting to a specific environment that would produce lack of generalization.

Figure ~\ref{fig:hand-bb} shows estimated bounding boxes over some test images once the Faster R-CNN network was trained.

\subsection{Hand Pose Regression}
\label{sec:hand-pose-regression}

The hand pose is computed by a modified ResNet50. The ResNet50 deep learning architecture is currently the state of the art CNN on image recognition, achieving a top-1 error ($22.85$) on the ImageNet validation split. This architecture introduces the 'residual' term, which consists of the aggregation of the input image to the output image of a convolution block. As a result, the output of a convolution block can be seen as the input image where the features activated by the filters are highlighted. In contrast, the output of a convolution layer in a default convolutional neural network is only the result of the neuron activation. If a neuron is not triggered on a certain region of the input image, the output remains with lower activation values. When the network computes the weights update in the backpropagation stage, the values on non-activated regions lead to very low upgrades, eventually even resulting in no upgrade at all, which causes the learning to stall. This issue is known as the vanishing gradient problem.
The inclusion of the 'residual' term helps fight the vanishing gradient problem and allows the design of even deeper architectures. Currently, the best performer on several tasks of the ImageNet challenge is based on the 'residual' approach introduced by ResNet.

As stated before, we used the ResNet50 version ($50$ layers). Our ResNet50 incarnation takes as input $224\times224$ pixels color images. The number of neurons from the last fully connected layer was adjusted to fit our problem. As there are $20$ joints to locate composed of $X$ and $Y$ values for each joint, the number of output neurons was set to $40$. We also changed the default softmax activation function was changed for a linear one.

This architecture was trained from scratch using the training split of the proposed dataset. First, hands were extracted using the bounding boxes, and the joint annotations were remapped accordingly and normalized in a range between $0$ and $1$. The network was trained on the resultant dataset for $200$ iterations reaching a training loss of $0.0000607571$ and a test loss of $0.00031547$. The test loss was used as an early stop criteria. The optimizer of choice was Adam with a learning rate of $0.001$. The loss function we used was mean squared error. 

\subsection{Experimentation}
\label{sec:experimentation}

In this section, we first describe the experimentation setup and then present results obtained for the proposed pipeline.

All timings and results for this pipeline were obtained by conducting the experiments in the following test setup: Intel Core i5-3570 with 8 GiB of Kingston HyperX 1600 MHz and CL10 DDR3 RAM on an Asus P8H77-M PRO motherboard (Intel H77 chipset). Principal storage was provided by a Seagate Desktop HDD. Additionally, the system included a NVIDIA GTX1080 GPU used for training and inference.

The execution of the full pipeline runs at $23$ frames per second, but it is affordable to run at $30-35$ frames per second with minor tweaks on the code. For example, the hand detector takes ~15 ms per frame on a NVIDIA GTX 1080. In the current system the hand detector is running for every frame. However,  with a naive motion estimation technique, bounding boxes could be tracked along the sequence and the hand position would only be recomputed when the system detects large displacements, thus reducing the computation time.

The frameworks of choice were Caffe 0.14-RC3 and Keras 1.2.0 with Tensor Flow 0.12 as the backbone, running on Ubuntu 14.04.02. Both were compiled using CMake 2.8.7, g++ 4.8.2, CUDA 8.0, and cuDNN v5.0. 

As described earlier, there are two trainable subsystem in our pipeline.
Both the hand detector and the hand pose estimator were trained on our dataset.

In order to validate our baseline approach, the full pipeline was processed using the test split, and achieved a validation mean absolute error of $0.06939\%$ over the input hand image ($224\times224)$, which means an error of $10$ pixels.

\begin{figure}[!hbt]
  \centering
  \includegraphics[width=0.32\textwidth, height=3.2cm]{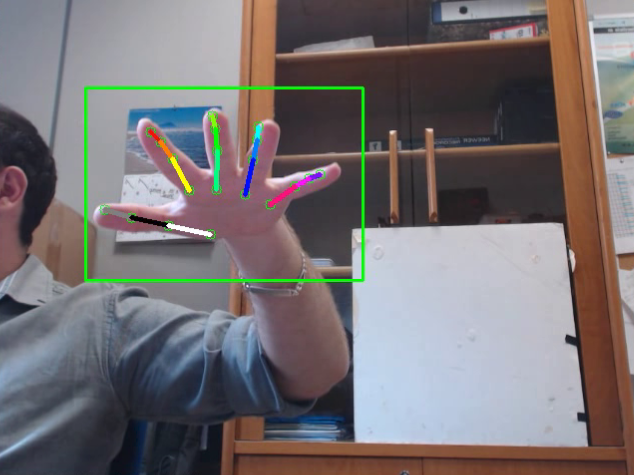}
  \vspace{1mm}
  \includegraphics[width=0.32\textwidth, height=3.2cm]{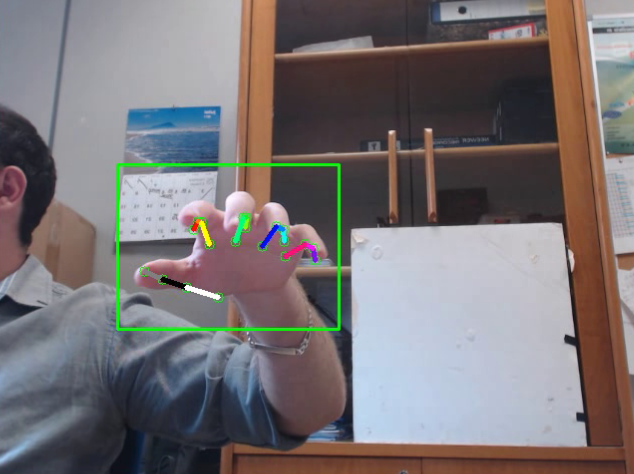}
  \includegraphics[width=0.32\textwidth, height=3.2cm]{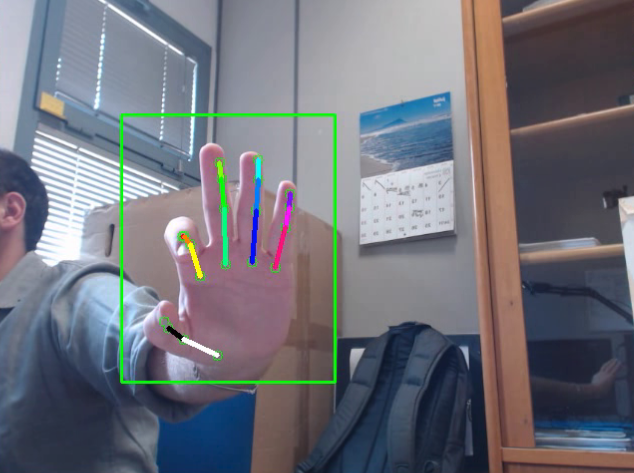}
  \vspace{1mm}
  \includegraphics[width=0.32\textwidth, height=3.2cm]{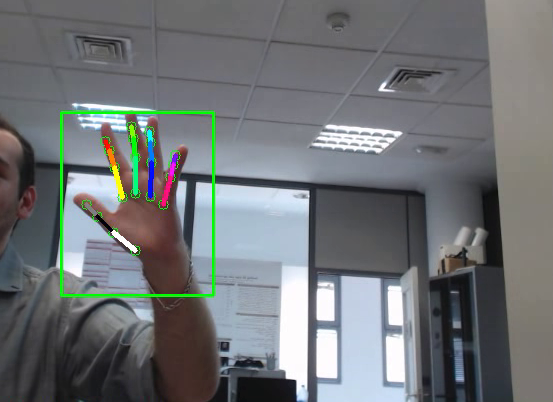}  		     \includegraphics[width=0.32\textwidth, height=3.2cm]{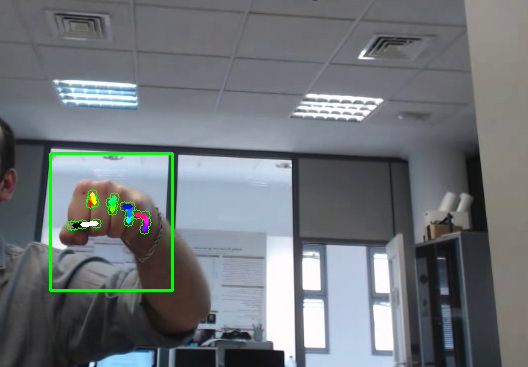}
  \includegraphics[width=0.32\textwidth, height=3.2cm]{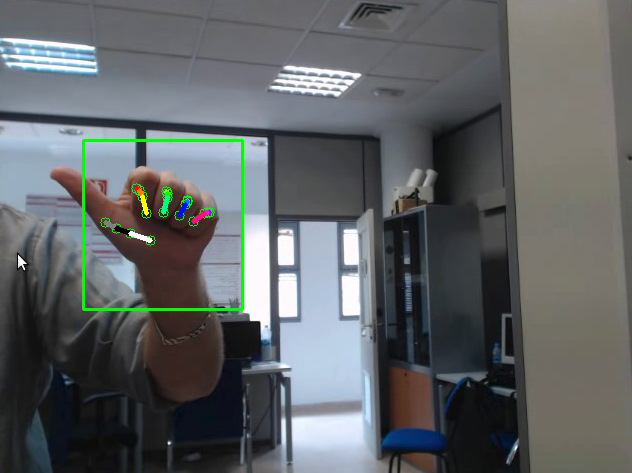}
  \caption{Our proposed system is able to accurately estimate hand joints position on a variety of hand poses. Nonetheless, it presents some failure cases as shown in the last image.}
  \label{fig:ok-poses}
\end{figure}

Figure \ref{fig:ok-poses} shows system performance for several hand poses. The last image shows a failure case due to the incorrect hand detection. As this approach is a dependent two-stage system, if the first step fails to detect the whole hand, the second system would try to regress the joints on wrong data, so it is very likely to fail.

\section{Conclusions}
\label{sec:conclusions}

Hand detection in both, 2D and 3D, is a challenging problem. In this work, we present a novel multiview hand pose dataset. It features over $20,500$ annotated color images of hands. The annotations include, besides several images taken from different perspectives, the bounding box and the 2D and 3D position of each joint in the hand.  Furthermore, we introduce a simple yet accurate architecture for real-time 2D hand pose estimation to be used as a baseline. This dataset could serve as a baseline for future works on hand detection. The dataset is available for download in \footnote{http://www.rovit.ua.es/dataset/mhpdataset/}.

Following on this work we plan to design a new architecture for robust 3D hand pose estimation in which the system will provide not only x and y coordinates in the image plane but x, y and z, and the normal of the palm on real world coordinates.

\section*{Acknowledgements}

This work has been funded by the Spanish Government TIN2016-76515-R grant for the COMBAHO project, supported with Feder funds.

\bibliographystyle{elsarticle-num} 
\bibliography{references}

\end{document}